\documentclass[epj]{webofc}
\usepackage[utf8]{inputenc}
\usepackage[varg]{txfonts}   
\usepackage{booktabs}
\usepackage{xcolor}
\definecolor{darkred}{rgb}{0.4,0.0,0.0}
\definecolor{darkgreen}{rgb}{0.0,0.4,0.0}
\definecolor{darkblue}{rgb}{0.0,0.0,0.4}
\usepackage[bookmarks,linktocpage,colorlinks,
    linkcolor = darkred,
    urlcolor  = darkblue,
    citecolor = darkgreen]{hyperref}
\usepackage{subfigure}
\usepackage{multicol}
\usepackage{bm}
\usepackage{caption}
\wocname{EPJ Web of Conferences}
\woctitle{Lattice2017}
\providecommand{\RN}[1]{%
  \textup{\uppercase\expandafter{\romannumeral#1}}%
}

\newcommand{\tsep}{\mathop{t_{\rm sep}}\nolimits}
\newcommand{\tsepi}{\mathop{t_{\rm sep} \to \infty}\nolimits}

\newcommand{\MeV}{\mathop{\rm MeV}\nolimits}
\newcommand{\GeV}{\mathop{\rm GeV}\nolimits}
\newcommand{\fm}{\mathop{\rm fm}\nolimits}

\graphicspath{{./figs/}}
\allowdisplaybreaks

\begin{document}

\selectlanguage{english}

\title{%
Isovector and flavor-diagonal charges of the nucleon
}
\author{%
  \firstname{Rajan} \lastname{Gupta}\inst{1}\fnsep\thanks{Speaker  \email{rajan@lanl.gov}. Los Alamos Report LA-UR-17-31334} \and
\firstname{Tanmoy} \lastname{Bhattacharya}\inst{1} \and
\firstname{Yong-Chull}  \lastname{Jang}\inst{1} \and
\firstname{Huey-Wen}  \lastname{Lin}\inst{3} \and
\firstname{Boram}  \lastname{Yoon}\inst{2}
}
\institute{%
Theoretical Division T-2, 
Los Alamos National Laboratory, 
Los Alamos, NM 87545, U.S.A.
\and
Computer, Computational, and Statistical Sciences CCS-7, 
Los Alamos National Laboratory,
Los Alamos, NM 87545, U.S.A.
\and
Department of Physics and Astronomy, 
Michigan State University, MI 48824, U.S.A
}

\abstract{

We present an update on the status of the calculations of isovector
and flavor-diagonal charges of the nucleon.  The calculations of the
isovector charges are being done using ten $2+1+1$-flavor HISQ
ensembles generated by the MILC collaboration covering the range of
lattice spacings $a \approx$ 0.12, 0.09, 0.06\,$\fm$ and pion masses
$M_\pi \approx$ 310, 220, 130\,$\MeV$.  Excited-states contamination
is controlled by using four-state fits to two-point correlators and
three-states fits to the three-point correlators.  The calculations of
the disconnected diagrams needed to estimate flavor-diagonal charges
are being done on a subset of six ensembles using the stocastic
method.  Final results are obtained using a simultaneous fit in
$M_\pi^2$, the lattice spacing $a$ and the finite volume parameter
$M_\pi L$ keeping only the leading order corrections. } \maketitle

\section{Introduction}\label{intro}

This talk presents an update on results given in
Refs.~\cite{Bhattacharya:2015wna,Bhattacharya:2015esa,Bhattacharya:2016zcn}
on isovector and flavor diagonal charges of the nucleon using our 
clover-on-HISQ lattice approach.  A summary of the $2+1+1$-flavor HISQ
ensembles generated by the MILC collaboration~\cite{Bazavov:2012xda}, and the number of
measurements made on them in the ongoing clover-on-HISQ study is given
in Table~\ref{tab:HISQ}. The improvements made since the results
reported in
Refs.~\cite{Bhattacharya:2015wna,Bhattacharya:2015esa,Bhattacharya:2016zcn}
are
\begin{itemize}
\item
cost-effective increase in statistics using the truncated solver
method and the coherent source sequential propagator technique. 
\item
The correction for possible bias in the truncated solver method is 
now made on all ensembles. 
\item
Addition of a second physical mass ensemble at weaker coupling, $a06m135$. 
\item
Excited-state contamination (ESC) is controlled using 4-states in the analysis of the 2-point 
correlation functions and 3-states for the 3-point functions. 
\item
Fits to 2- and 3-point functions are done using the full covariance matrix in the mimimization of $\chi^2$. \looseness-1
\item
A simultaneous fit in $a$, $M_\pi$ and $M_\pi L$ is used to extract
physical results in the limits $a\to 0$, $M_\pi=135$~MeV and $M_\pi L
\to \infty$ from lattice data obtained at different values of $a$,
$M_\pi$ and $M_\pi L$.
\end{itemize}
Associated results for the isovector form factors, 
$G_A(Q^2)$, $\tilde{G}_P(Q^2)$, $G_E(q^2)$ and $G_M(q^2)$, on these ensembles were presented by
Yong-Chull Jang at this conference~\cite{Jang:2017}.

\begin{table}
  \renewcommand{\arraystretch}{1.2} 
  \resizebox{0.98\linewidth}{!}{
    \begin{tabular}{l|ccc|cc|ccc}
    \hline
      Ensemble ID & $a$ (fm) & $M_\pi^{\rm sea}$ (MeV) & $M_\pi^{\rm val}$ (MeV) & $L^3\times T$    & $M_\pi^{\rm val} L$ & $N_\text{conf}$  & $N_{\rm meas}^{\rm HP}$  & $N_{\rm meas}^{\rm AMA}$  \\
  \hline
  $a12m310 $   & 0.1207(11) & 305.3(4) & 310(3) & $24^3\times 64$ & 4.55 & 1013 & 8104  &   64,832   \\
  $a12m220S$   & 0.1202(12) & 218.1(4) & 225(2) & $24^3\times 64$ & 3.29 & 946  & 3784  &   60,544   \\
  $a12m220 $   & 0.1184(10) & 216.9(2) & 228(2) & $32^3\times 64$ & 4.38 & 744  & 2976  &   47,616   \\
  $a12m220L$   & 0.1189(9)  & 217.0(2) & 228(2) & $40^3\times 64$ & 5.49 & 1010 & 8080  &   68,680   \\
      \hline                                      
  $a09m310 $   & 0.0888(8)  & 312.7(6) & 313(3) & $32^3\times 96$ & 4.51 & 2264 & 9056  &  114,896   \\
  $a09m220 $   & 0.0872(7)  & 220.3(2) & 226(2) & $48^3\times 96$ & 4.79 & 964  & 3856  &  123,392   \\
  $a09m130 $   & 0.0871(6)  & 128.2(1) & 138(1) & $64^3\times 96$ & 3.90 & 883  & 7064  &  84,768    \\
      \hline                                      
  $a06m310 $   & 0.0582(4)  & 319.3(5) & 320(2) & $48^3\times 144$& 4.52 & 1000 & 8000  &  64,000    \\
$a06m310^\ast$ &            &          &        &                 &      & 500  & 2000  &  64,000    \\
  $a06m220$    & 0.0578(4)  & 229.2(4) & 235(2) & $64^3\times 144$& 4.41 & 650  & 2600  &  41,600    \\
$a06m220^\ast$ &            &          &        &                 &      & 650  & 2600  &  41,600    \\
  $a06m135 $   & 0.0568(1)  & 135.5(2) & 136(2) & $96^3\times 192$& 3.74 & 322  & 1288  &  20,608    \\
      \hline
    \end{tabular}
  }
\caption{Summary of the 2+1+1-flavor HISQ ensembles generated by the
  MILC Collaboration~\protect\cite{Bazavov:2012xda} and used in our
  clover-on-HISQ study. The $a06m310^\ast$ and the $a06m220^\ast$
  ensembles represent a second analysis with larger source smearings,
  $\sigma=12$ and $11$, respectively, as described in
  Ref.~\protect\cite{Bhattacharya:2016zcn}. }
\label{tab:HISQ}
\end{table}

\section{Controlling excited-state contamination} 

Our goal is to extract the matrix elements of various bilinear quark
operators between ground state nucleons. The lattice operator $\chi(x)
= \epsilon^{abc} \left[ {q_1^a}^T(x) C \gamma_5 \frac{(1 \pm
    \gamma_4)}{2} q_2^b(x) \right] q_1^c(x)$ used to create and
annihilate the nucleon state couples to the nucleon, all its
excitations and multiparticle states with the same quantum
numbers. The correlation functions, therefore, get contributions from
all these intermediate states. This ESC can
be evaluated and controlled using fits including as many states as the
data allow in the spectral decomposition of the two- and three-point
functions.  In our study we use: \looseness-1
\begin{equation}
C^\text{2pt}
  (t_f,t_i) = 
  {|{\cal A}_0|}^2 e^{-aM_0 (t_f-t_i)} + {|{\cal A}_1|}^2 e^{-aM_1 (t_f-t_i)} + 
  {|{\cal A}_2|}^2 e^{-aM_2 (t_f-t_i)} + {|{\cal A}_3|}^2 e^{-aM_3 (t_f-t_i)} + \ldots \,, 
\label{eq:2pt}
\end{equation}
\begin{align}
C^\text{3pt}_{\Gamma}(t_f,t,t_i) &= 
    |{\cal A}_0|^2 \langle 0 | \mathcal{O}_\Gamma | 0 \rangle  e^{-aM_0 (t_f - t_i)} +{}
    |{\cal A}_1|^2 \langle 1 | \mathcal{O}_\Gamma | 1 \rangle  e^{-aM_1 (t_f - t_i)} +{}
    |{\cal A}_2|^2 \langle 2 | \mathcal{O}_\Gamma | 2 \rangle  e^{-aM_2 (t_f - t_i)} +{}\nonumber\\
  & {\cal A}_1{\cal A}_0^* \langle 1 | \mathcal{O}_\Gamma | 0 \rangle  e^{-aM_1 (t_f-t)} e^{-aM_0 (t-t_i)} +{}
    {\cal A}_0{\cal A}_1^* \langle 0 | \mathcal{O}_\Gamma | 1 \rangle  e^{-aM_0 (t_f-t)} e^{-aM_1 (t-t_i)} +{}\nonumber\\
  & {\cal A}_2{\cal A}_0^* \langle 2 | \mathcal{O}_\Gamma | 0 \rangle  e^{-aM_2 (t_f-t)} e^{-aM_0 (t-t_i)} +{}
    {\cal A}_0{\cal A}_2^* \langle 0 | \mathcal{O}_\Gamma | 2 \rangle  e^{-aM_0 (t_f-t)} e^{-aM_2 (t-t_i)} +{}\nonumber\\
  & {\cal A}_1{\cal A}_2^* \langle 1 | \mathcal{O}_\Gamma | 2 \rangle  e^{-aM_1 (t_f-t)} e^{-aM_2 (t-t_i)} +{}
    {\cal A}_2{\cal A}_1^* \langle 2 | \mathcal{O}_\Gamma | 1 \rangle  e^{-aM_2 (t_f-t)} e^{-aM_1 (t-t_i)} + \ldots \,,
\label{eq:3pt}
\end{align}
where we have shown all contributions from the ground state
$|0\rangle$ and the first three excited states $|1\rangle$,
$|2\rangle$ and $|3\rangle$ with masses $M_1$, $M_2$ and $M_3$ to the
two-point functions, and from the first two excited states for the
three-point functions.  The analysis, using Eqs.~\eqref{eq:2pt}
and~\eqref{eq:3pt}, is called a ``3-state fit'' or ``4-state fit''
depending on the number of intermediate states included.  The 2-state
analysis (keeping one excited state) of 3-point functions requires
extracting seven parameters ($M_0$, $M_1$, ${\cal A}_0$, ${\cal A}_1$,
$\langle 0 | \mathcal{O}_\Gamma | 0 \rangle$, $ \langle 1 |
\mathcal{O}_\Gamma | 0 \rangle$ and $ \langle 1 | \mathcal{O}_\Gamma |
1 \rangle$) from fits to the two- and three-point functions. The
3-state analysis introduces five additional parameters: $M_2$, ${\cal
  A}_2$, $\langle 0 | \mathcal{O}_\Gamma | 2 \rangle$, $\langle 1 |
\mathcal{O}_\Gamma | 2 \rangle$ and $\langle 2 | \mathcal{O}_\Gamma |
2 \rangle$.  On each ensemble we generate data at multiple values of
$\tsep \equiv t_f-t_i \equiv \tau$.  A simultaneous fits to the data
at all $\tau$ and $t$ allows us to extract the charges in the limit
$\tsepi$, i.e., the ground state matrix element $\langle 0 |
\mathcal{O}_\Gamma | 0 \rangle$. Throughout this paper, values of $t$
and $\tau = \tsep$ are in lattice units unless explicitly stated.

\begin{figure*}[tb]
\subfigure{
    \includegraphics[width=0.32\linewidth]{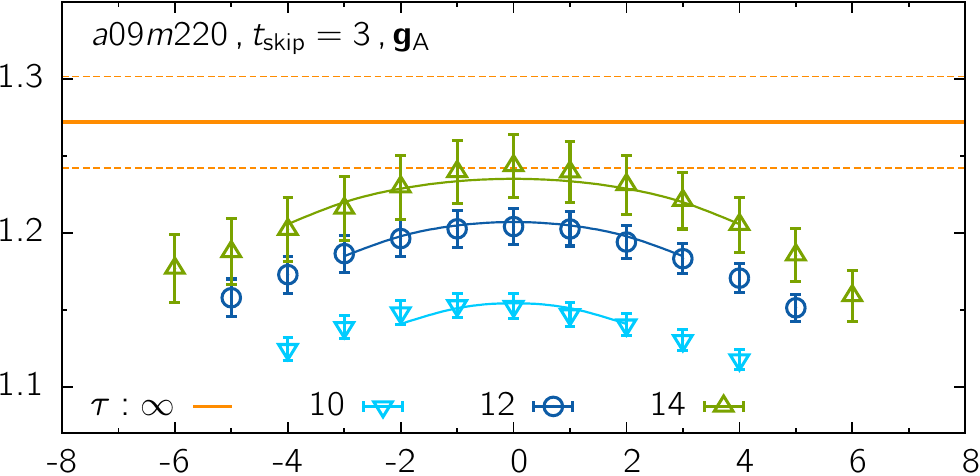}
    \includegraphics[width=0.32\linewidth]{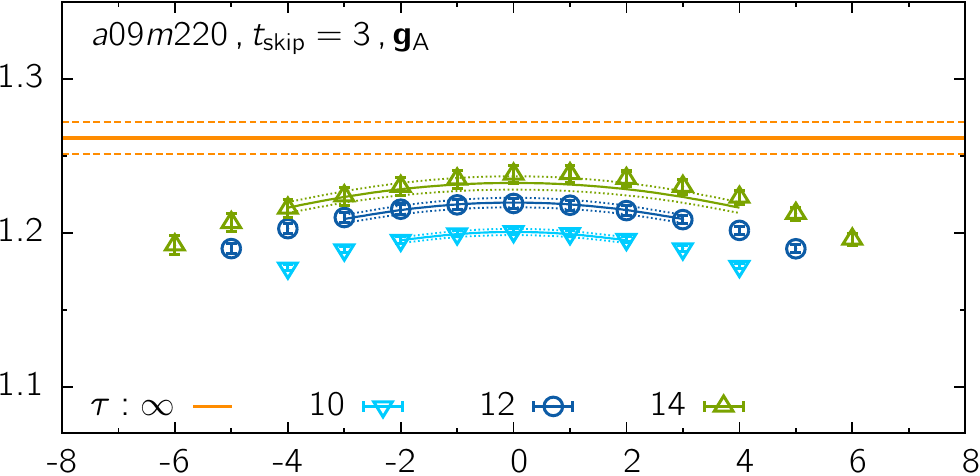}
    \includegraphics[width=0.32\linewidth]{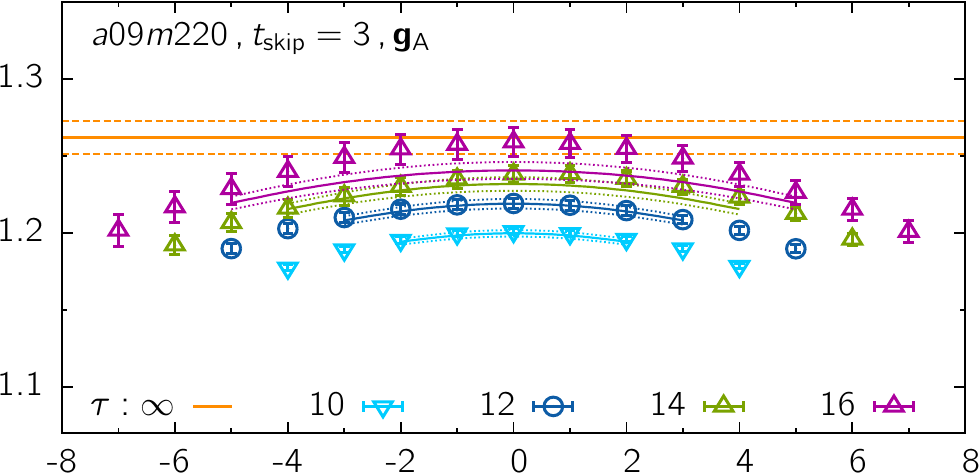}
}
\subfigure{
    \includegraphics[width=0.32\linewidth]{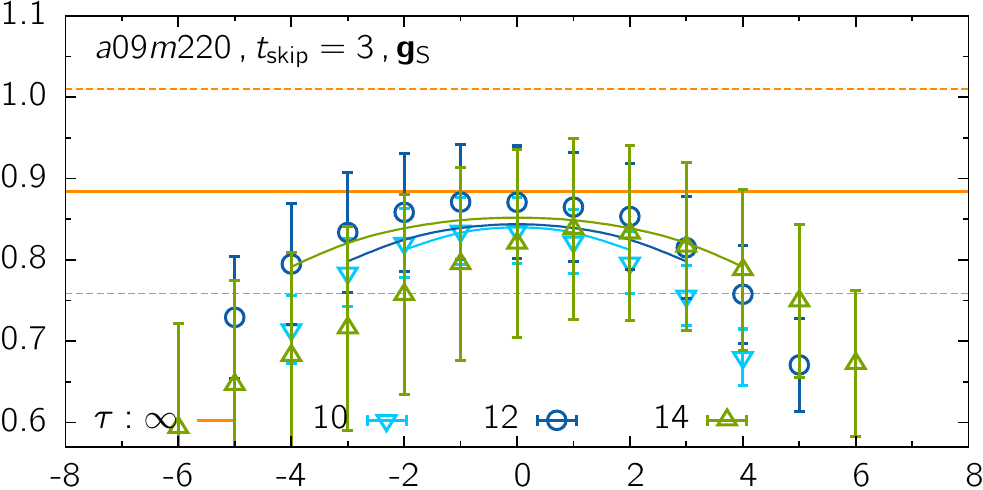}
    \includegraphics[width=0.32\linewidth]{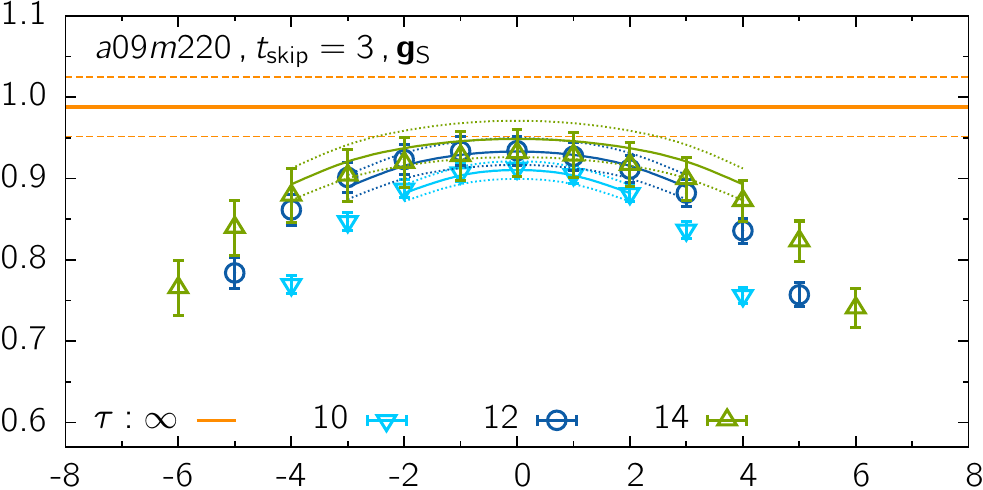}
    \includegraphics[width=0.32\linewidth]{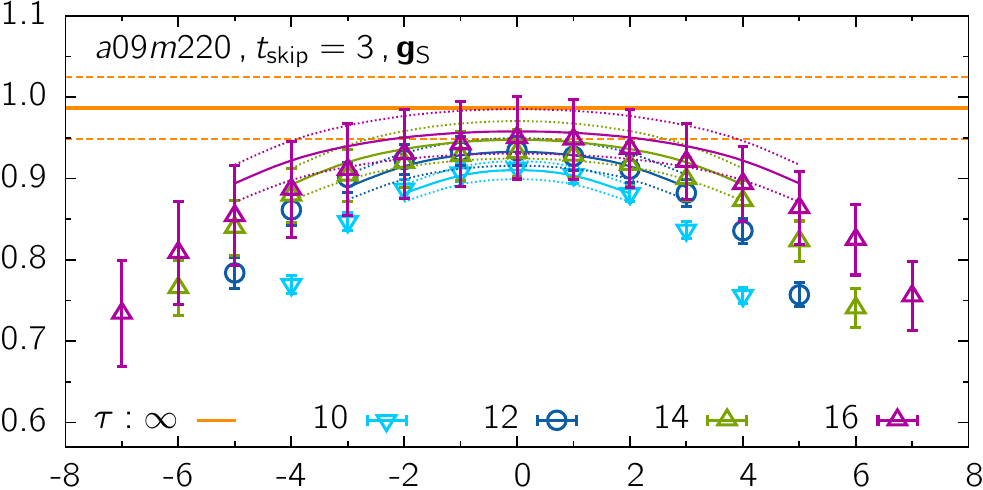}
}
\subfigure{
    \includegraphics[width=0.32\linewidth]{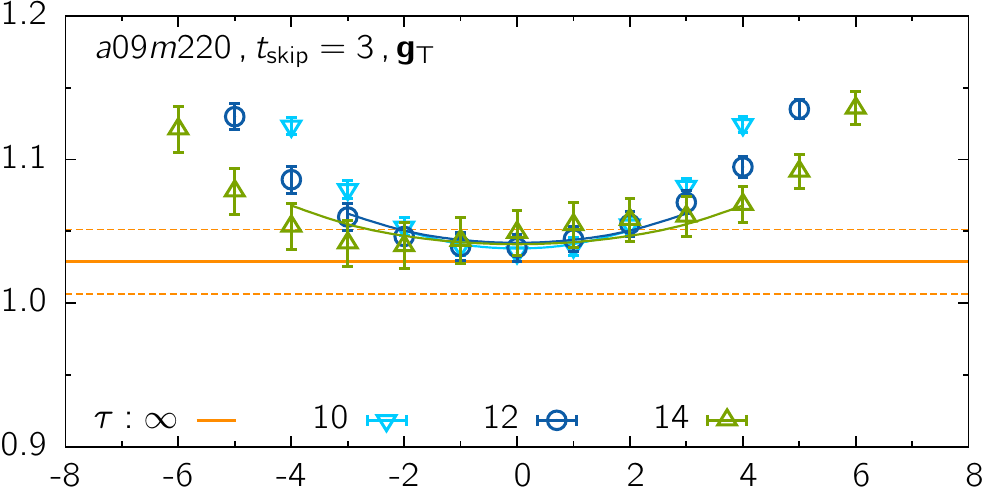}
    \includegraphics[width=0.32\linewidth]{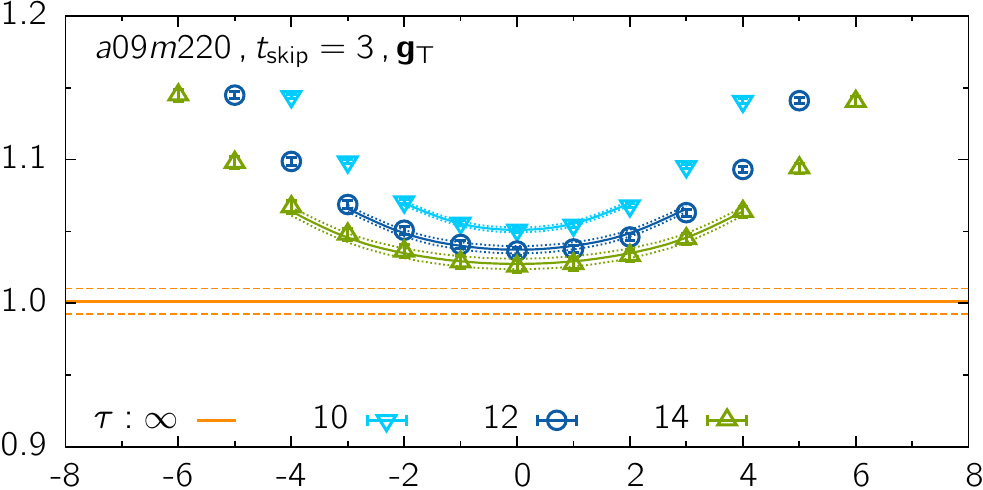}
    \includegraphics[width=0.32\linewidth]{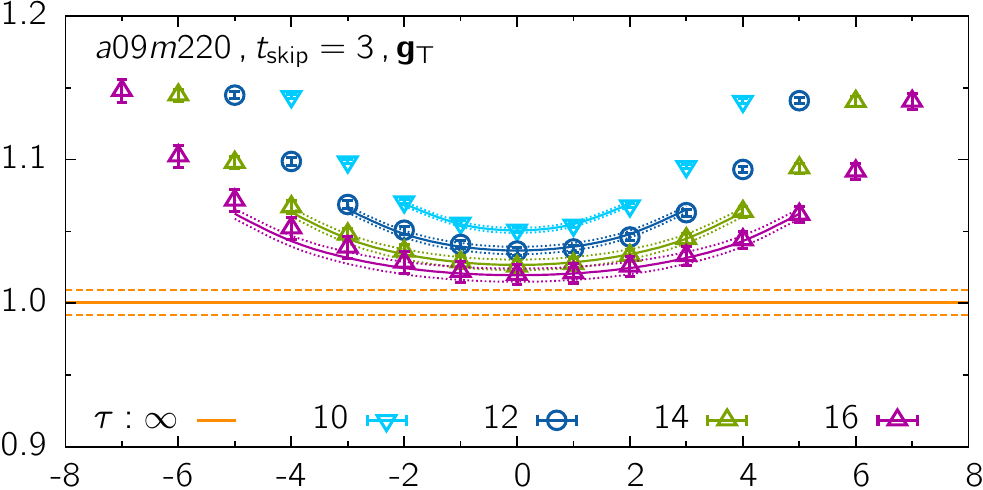}
}
\caption{Illustration of the control over ESC in the
  isovector charges, $g_A^{u-d}$, $g_S^{u-d}$ and $g_T^{u-d}$, with
  higher statistics on the $a09m220$ ensemble. The left panels give
  the results based on 8000 HP measurements reported in
  Ref.~\protect\cite{Bhattacharya:2016zcn}. The middle and right
  panels show new results with 123,392 AMA measurements.  While the
  results from all three fits are consistent, the reliability of the fits,  
  especially for  $g_S^{u-d}$, is greatly improved when (i) the monotonic
  convergence in $\tau$ is manifest, and (ii) the fits and the values without 
  (middle panels) and with (right panels) the $\tsep=16$ data overlap. }
  \label{fig:ESC220}
\end{figure*}

Fig.~\ref{fig:ESC220} shows data from the $a09m220$ ensemble and
highlights a number of features in the data and control over ESC using
the simultaneous fit in $t$ and $\tau$: (i) with increased statistical
precision (HP $\to$ AMA), the convergence w.r.t. $\tau$ is demonstrated to be 
monotonic in all three charges, $g_{A,S,T}$.  Previous HP estimates
for both $g_{S,T}$ were affected by a lack thereof. In fact, we now
require this monotonic behavior when evaluating the statistical
reliablity of data. (ii) Increasing the source smearing size
$\sigma=5.5 \to 7.0$ reduced ESC in $g_{A,S}$, but marginally
increases it in $g_T$.  (iii) The fits including $\tau=16$ data (right
panels) confirm the results of the fits without it (middle panels),
indicating convergence.

The renormalized values of the isovector charges, using the
renormalization factors given in Ref.~\cite{Bhattacharya:2016zcn}, are
summarized in Table~\ref{tab:charges}.  The table also reproduces the
CalLat results for $g_A^{u-d}$ from Ref.~\cite{Berkowitz:2017gql} on the
five ensembles analyzed by both Collaborations. We compare these
results in Sec.~\ref{sec:CalLat}.

\begin{table}
\renewcommand{\arraystretch}{1.2} 
\resizebox{0.98\linewidth}{!}{
\begin{tabular}{l|cccc|ccc|ccc}
\hline
Charge        & $a12m310$  & $a12m220S$  & $a12m220$  & $a12m220L$  & $a09m310$  & $a09m220$  & $a09m130$  & $a06m310$  & $a06m220$  & $a06m130$  \\
\hline
$g_A^{u-d}$   & 1.251(19)  & 1.223(44)   & 1.238(24)  & 1.264(20)   & 1.217(14)  & 1.239(17)  & 1.245(32)  & 1.209(28)  & 1.206(21)  & 1.213(37)  \\
              &            &             &            &             &            &            &            & 1.205(24)  & 1.240(26)  &            \\
\hline
$g_A^{u-d}$ CL& 1.237(07)  & 1.272(28)   & 1.259(15)  & 1.252(21)   & 1.258(14)  &            &            &            &            &            \\
\hline
$g_S^{u-d}$   & 0.840(54)  & 0.902(253)  & 0.952(99)  & 0.742(53)   & 0.919(51)  & 0.896(56)  & 0.926(128) & 1.110(90)  & 0.978(74)  & 0.943(188)  \\
              &            &             &            &             &            &            &            & 0.970(78)  &            &            \\
\hline
$g_T^{u-d}$   & 1.035(37)  & 1.009(53)   & 1.021(38)  & 1.003(38)   & 1.043(29)  & 1.011(29)  & 0.969(35)  & 1.015(30)  & 1.022(27)  & 1.029(36)  \\
              &            &             &            &             &            &            &            & 1.037(30)  & 1.018(34)  &            \\
\hline
\end{tabular}
}
\caption{Results for the renormalized isovector charges in the
  $\overline{\rm MS}$ scheme at 2~GeV. In the third row we reproduce
  CalLat's results (labeled CL) for
  $g_A^{u-d}$ from Ref.~\protect\cite{Berkowitz:2017gql} on the five HISQ
  ensembles analyzed by both
  collaborations~\protect\cite{Berkowitz:2017gql}.}
\label{tab:charges}
\end{table}

%

%

\section{Simultaneous fit in $a$, $M_\pi$ and $M_\pi L$} 

Having calculated renormalized charges at various values of $a$,
$M_\pi$ and $M_\pi L$, we perform a simultaneous fit to obtain results
in the limit $a \to 0$, $M_\pi=135$~MeV and $M_\pi L \to \infty$. When
fitting data given in Table~\ref{tab:charges} from the 10 HISQ
ensembles, we include only the lowest order correction terms
~\cite{Bhattacharya:2016zcn}:
\begin{equation}
  g_{A,S,T}^{u-d} (a,M_\pi,L) = c_1 + c_2a + c_3 M_\pi^2 +  c_4 M_\pi^2 {e^{-M_\pi L}} \,.
\label{eq:CextrapgAST} 
\end{equation}
Eq.~\eqref{eq:CextrapgAST} corrects a mistake made in
Ref.~\cite{Bhattacharya:2016zcn} for the analysis of the isovector
$g_S^{u-d}$.  The leading chiral term is proportional to $M_\pi^2$ for the
isovector case, and proportional to $M_\pi$ for the flavor diagonal
cases.

\begin{figure*}[tb]
\subfigure{
    \includegraphics[width=0.32\linewidth]{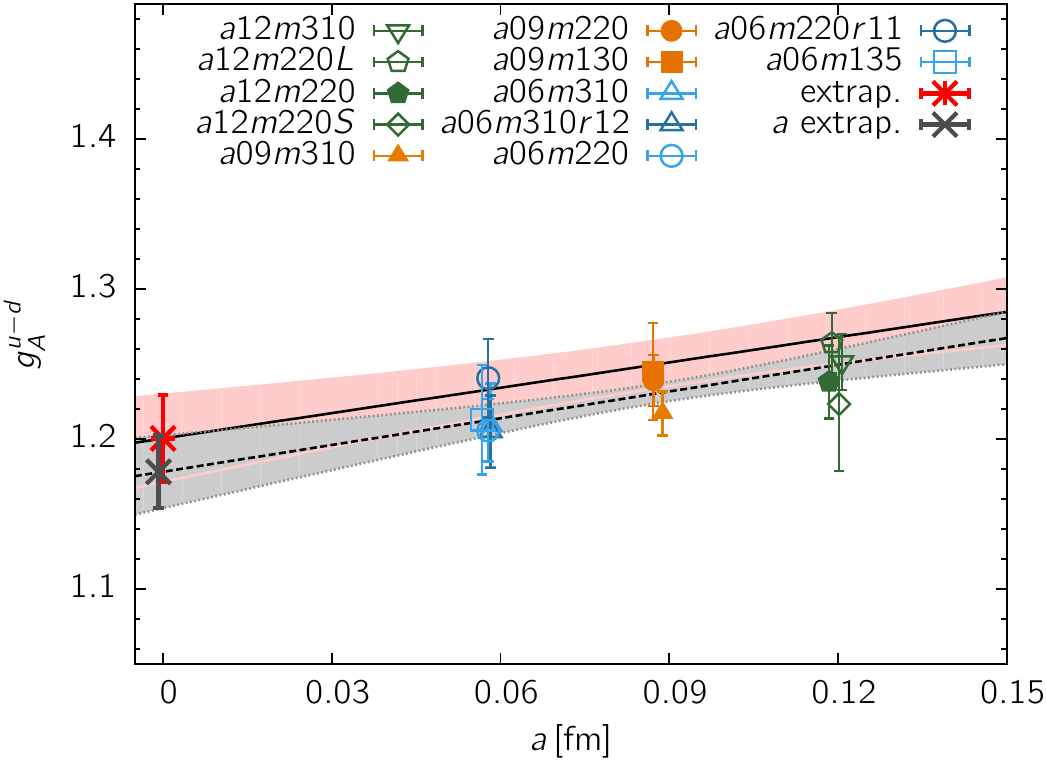}
    \includegraphics[width=0.32\linewidth]{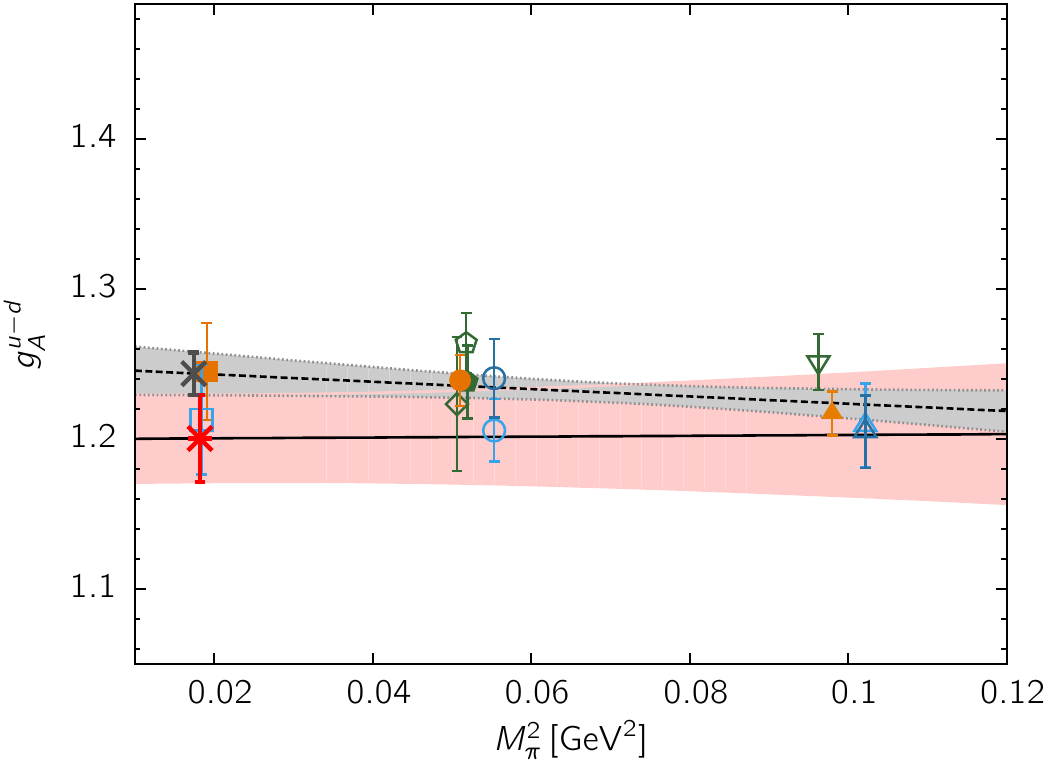}
    \includegraphics[width=0.32\linewidth]{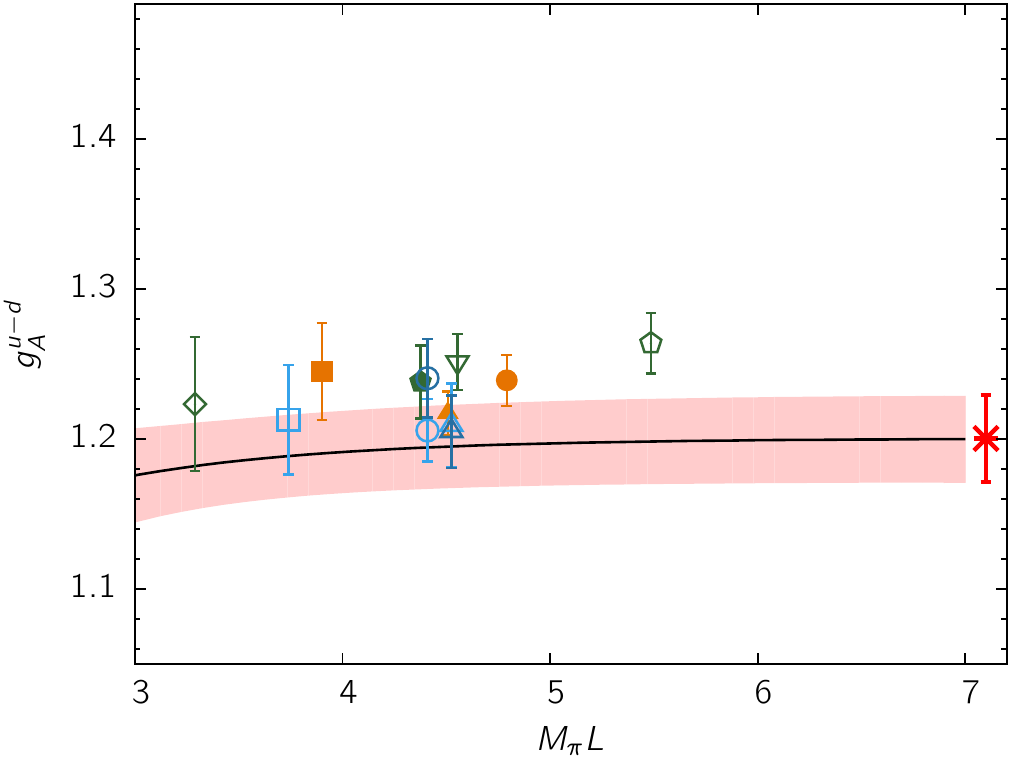}
}
\subfigure{
    \includegraphics[width=0.32\linewidth]{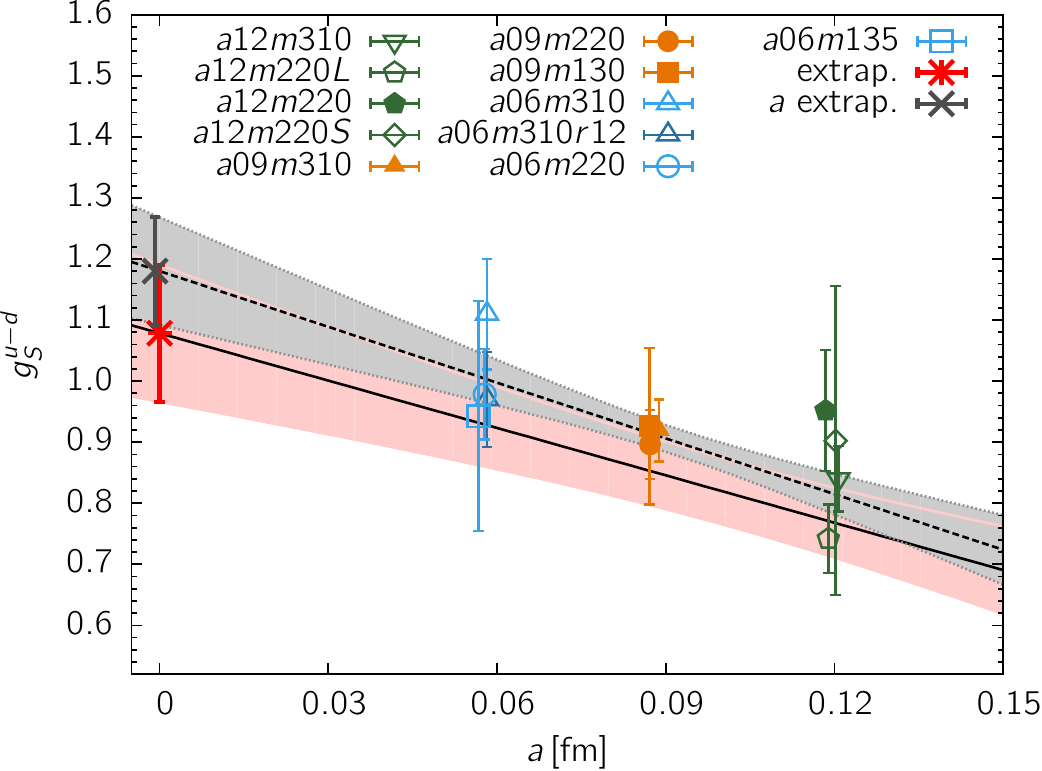}
    \includegraphics[width=0.32\linewidth]{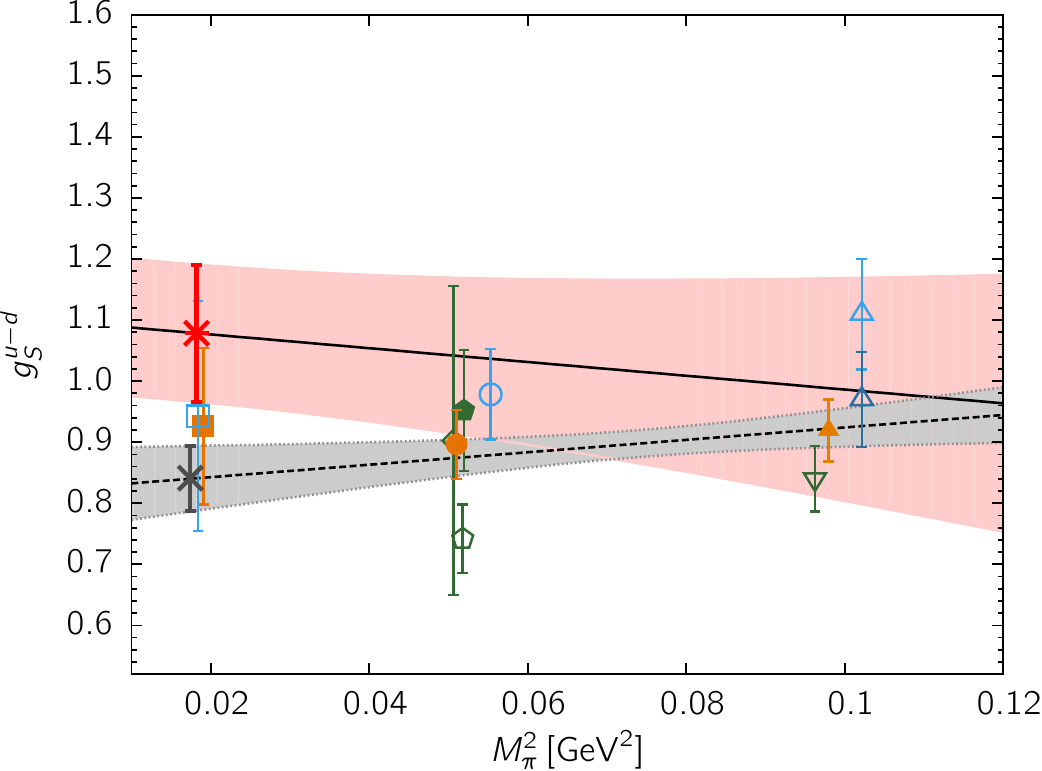}
    \includegraphics[width=0.32\linewidth]{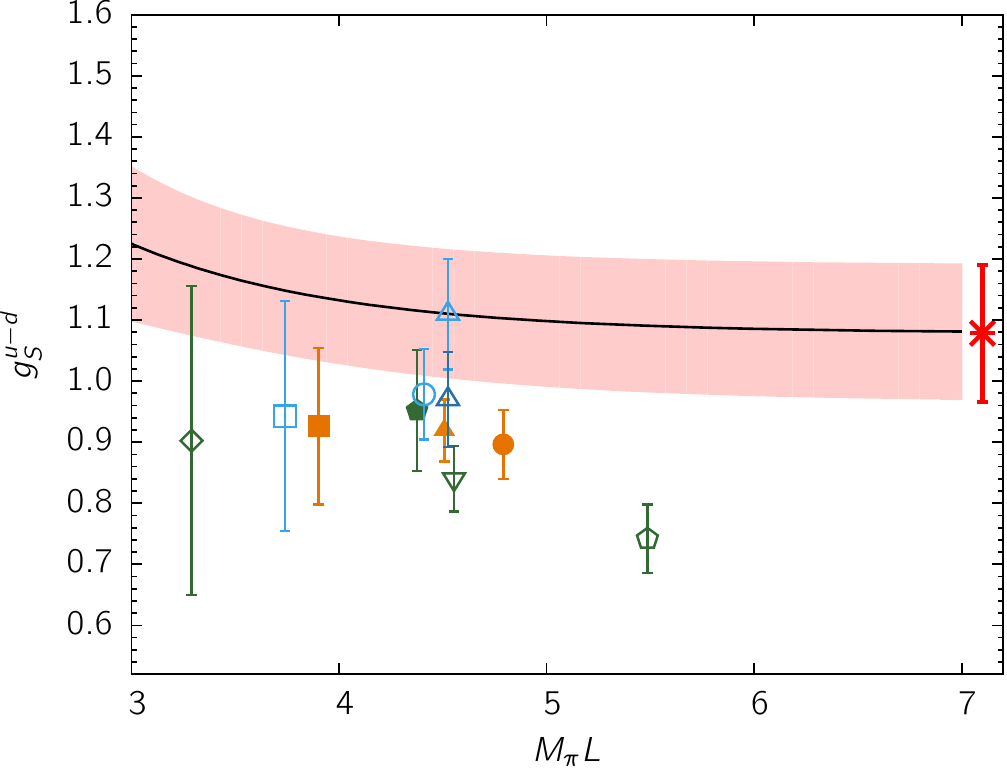}
}
\subfigure{
    \includegraphics[width=0.32\linewidth]{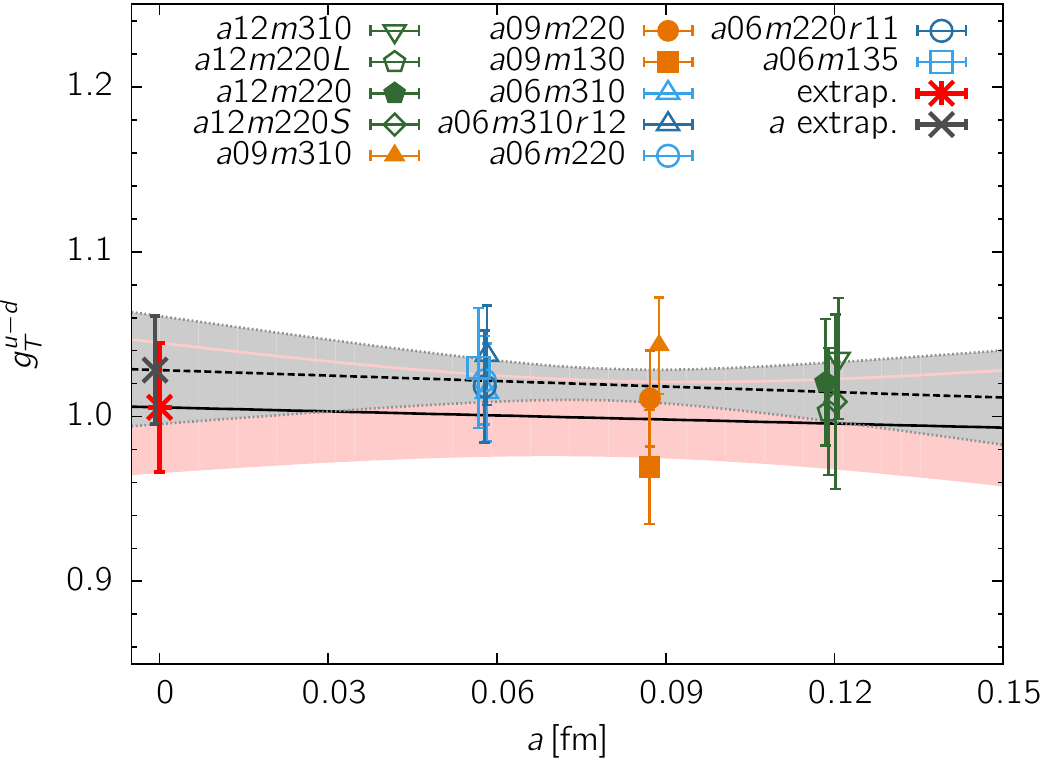}
    \includegraphics[width=0.32\linewidth]{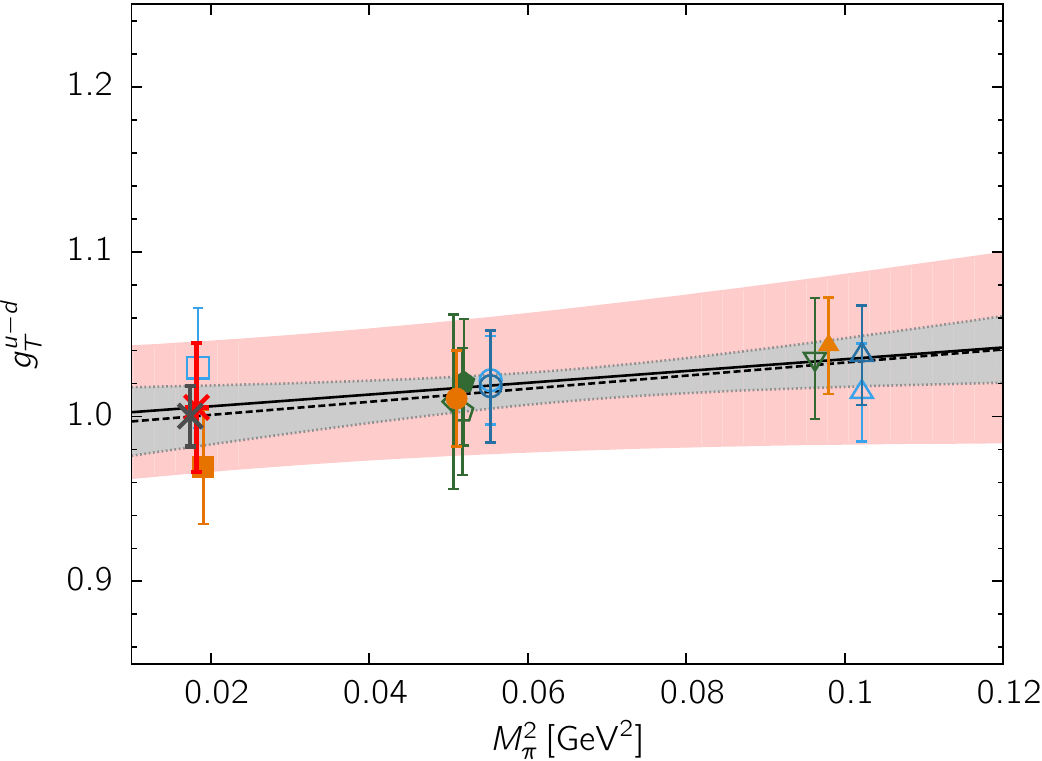}
    \includegraphics[width=0.32\linewidth]{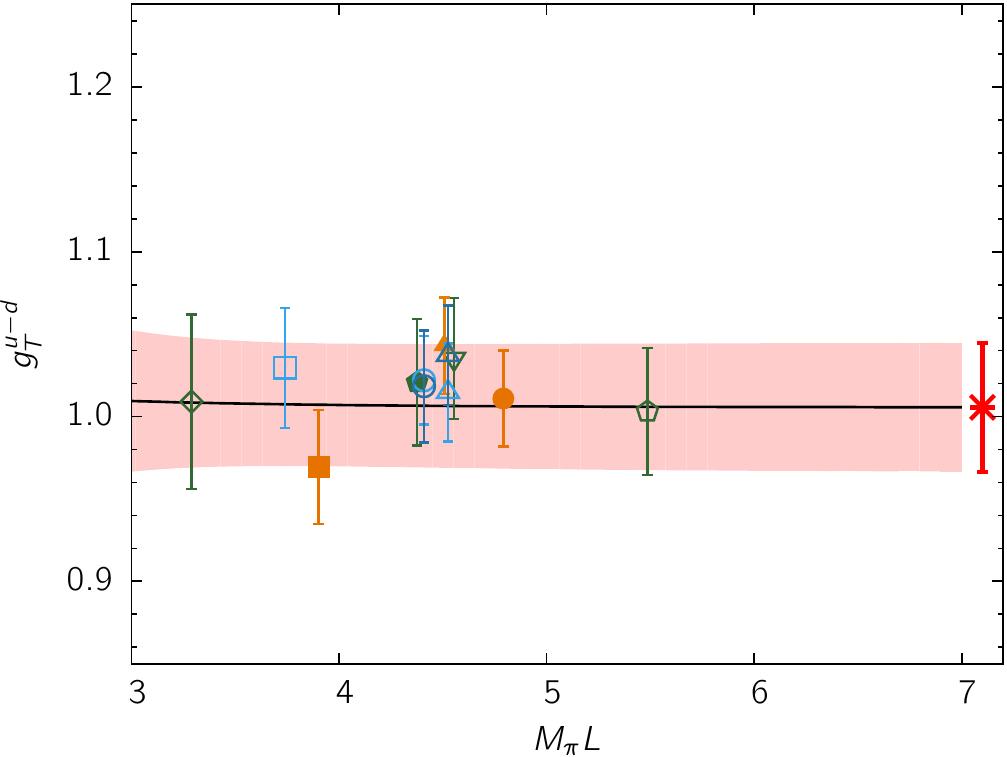}
}
\caption{The 12-point fit using Eq.~\protect\eqref{eq:CextrapgAST} to
  the data for the renormalized isovector charges $g_A^{u-d}$ and
  $g_T^{u-d}$ in the $\overline{{\rm MS}}$ scheme at $2\GeV$, and the
  11-point fit for $g_S^{u-d}$.  (The $a06m220^\ast$ point is
  neglected as the 3-point data for $g_S^{u-d}$ are not monotonic in
  $\tau$.) The result of the simultaneous extrapolation to the
  physical point defined by $a\rightarrow 0$, $M_\pi \rightarrow
  M_{\pi^0}^{{\rm phys}}=135$~MeV and $L \rightarrow \infty$ are
  marked by a red star.  The error bands in each panel are the result
  of a simultaneous fit but shown as a function of a single variable
  in the three panels.  The overlay in the left (middle) panels with
  the dashed line within the grey band, is the fit to the data versus
  $a$ ($M_\pi^2$), i.e., neglecting dependence on the other two
  variables.}
  \label{fig:conUmD_extrap12}
\end{figure*}

Fig.~\ref{fig:conUmD_extrap12} shows that with reduced errors due to
higher statistics data from 4 ensembles ($a12m220S$, $a12m220$,
$a09m220$ and $a09m310$) and the addition of the second physical-mass
ensemble $a06m135$, the behavior versus $a$, $M_\pi$ and $M_\pi L$ in
the simultaneous fits is visibly clearer compared to the ``9-point''
fits presented in Ref.~\cite{Bhattacharya:2016zcn}.  There is no
significant evidence for finite volume corrections in any of the three
charges for $M_\pi L > 3.5$. There is some dependence of $g_S^{u-d}$
on $M_\pi^2$. The most evident trends are the positive slope versus
$a$ in $g_A^{u-d}$ and the negative slope versus $a$ in $g_S^{u-d}$.
Based on these fits shown in Fig.~\ref{fig:conUmD_extrap12} and made using
Eq.~\eqref{eq:CextrapgAST}, our final estimates for the isovector
charges, in the $\overline{\rm MS}$ scheme at 2~GeV, are:\looseness-1
\begin{align}
  g_A^{u-d}  &= 1.20(3) \,, \\
  g_S^{u-d}  &= 1.08(11) \,, \\
  g_T^{u-d}  &= 1.01(4) \,.
  \label{eq:gFinal}
\end{align}
Given the improved data and the fits in
Fig.~\ref{fig:conUmD_extrap12}, the continued $2.5\sigma$ deviation of
$g_A^{u-d}$ from the experimental value indicates that we are
underestimating our errors. The largest change from results presented
in Ref.~\cite{Bhattacharya:2016zcn} is the $1\sigma$ increase in the 
estimate of $g_S^{u-d}$. Most of this increase is due to correcting
the form of the leading chiral term, i.e., $M_\pi \to M_\pi^2$, in
Eq.~\eqref{eq:CextrapgAST}.  The major source of error in $g_T^{u-d}$ is
now from the renormalization factor due to the poor convergence of the
perturbative matching between the $\overline{\rm MS}$ and RI-sMOM
schemes.

\section{Comparison with CalLat Results for $g_A^{u-d}$}
\label{sec:CalLat}

It is important to understand why our result for $g_A^{u-d} =1.20(3)$
presented in Eq.~\eqref{eq:gFinal} differs from a similarly precise
CalLat result $g_A^{u-d} = 1.278(21)(26)$ that agrees with the
experimental value $g_A^{u-d} = 1.276(2)$ when the data shown in
Table~\ref{tab:charges} on the 5 common ensembles are consistent.  Our
conclusion is that the majority of the difference comes from the final
extrapolation in $a$. While we find a positive slope controlled  by the
data on the three $a=0.06$~fm ensembles, CalLat finds a negative slope
anchored by the data on the coarser lattices. So the question is
whether the differences in the two methods are manifest only at weaker
couplng or are there systematic effects being missed in one or both
calculations?

The two sets of calculations are being done on the same 2+1+1-flavor
HISQ ensembles, but there are notable differences. These include: (i)
M\"obius domain wall versus clover for the valence quark action; (ii)
gradient flow smearing with $t_{gf}/a=1$ versus one HYP smearing to
smooth the lattices; (iii) different construction of the sequential
propagator. CalLat inserts a zero-momentum projected axial current in all timeslices on the
lattice simultaneously. This gives a summed contribution from all
timeslices between and on the source and sink points plus all timeslices 
outside. CalLat thus uses a 2-state fit to $g_A = C_3(\tau+1)/C_2(\tau+1) -
C_3(\tau)/C_2(\tau)$ to extract the charge where $C_3$ are 3-point
functions with the insertion on {\it all} timeslices; (iv) CalLat 
report a much better statistical signal with fewer measurements. 

The better statistical precision of the CalLat results for a given
number of measurements is easy to understand: the CalLat fits to
extract $g_A^{u-d}$ are based on a range of $\tau$ values that is
shifted by 6--8 timeslices to smaller $\tau$ compared to our fit
range. Since the errors in the data increase by a factor of two for
every increase in $\tau$ by two lattice units, they gain a factor of
up to $2^4$. Choosing values of $\tau$ within the range we have
simulated, our estimates for the quantity they calculate, $g_A =
C_3(\tau+1)/C_2(\tau+1) - C_3(\tau)/C_2(\tau)$, have similar
errors. Note, also, that the CPU cost of the CalLat calculation is,
ensemble by ensemble, higher because they simulate domain wall
fermions and did not use the multigrid algorithm for propagator
inversion.

The question, therefore, reduces to why their data can be fit starting
at much smaller values of $\tau$?  The correction due to ESC in their
smeared-smeared data is less than $10\%$ even at $\tau \sim 3$ on the
five common ensembles.  The necessary condition to achieve this in our
approach is reducing the overlap of the nucleon interpolating operator
with the excited/multiparticle states to essentially zero.  Since the
source smearing used by the two collaborations is similar and the
neutron interpolating operator is the same, the difference ``must''
come from the use of the gradient flow to smear the lattices. Further
investigations are needed to confirm this interpretation (similar
source smearing on gradient flow smoothed lattices produces sources
with much smaller overlap with excited states) since one does not,
{\it a priori}, expect the gradient flow smoothed lattices to change
the overlap with the excited states, but only to reduce ultraviolet
fluctuations.

\section{Disconnected Contributions}
\label{sec:disc}

\begin{figure*}[tb]
\subfigure{
\includegraphics[height=0.90in,trim={0.05cm  0.02cm 0 0},clip]{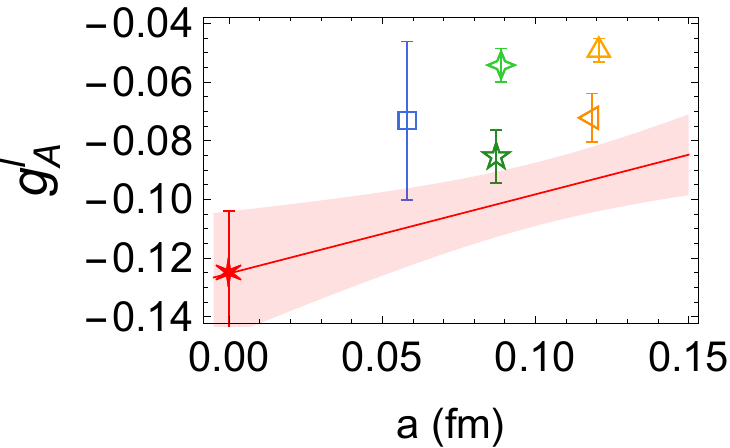}
\includegraphics[height=0.90in,trim={1.00cm  0.02cm 0 0},clip]{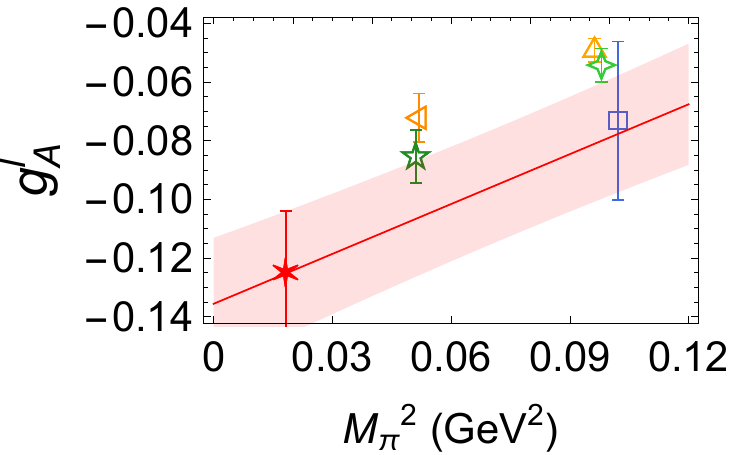}
\includegraphics[height=0.90in,trim={0.05cm  0.02cm 0 0},clip]{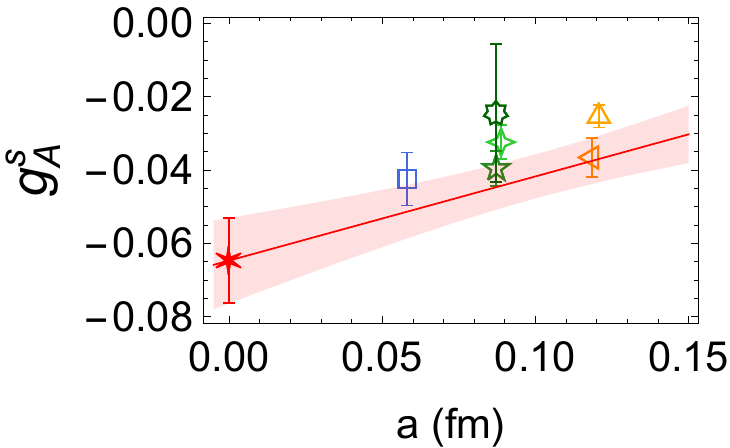}
\includegraphics[height=0.90in,trim={1.00cm  0.02cm 0 0},clip]{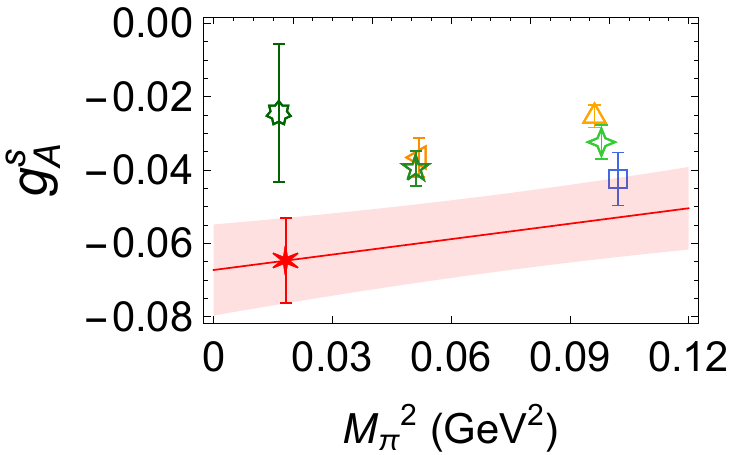}
}
\subfigure{
\includegraphics[height=0.96in,trim={0.05cm  0.02cm 0 0},clip]{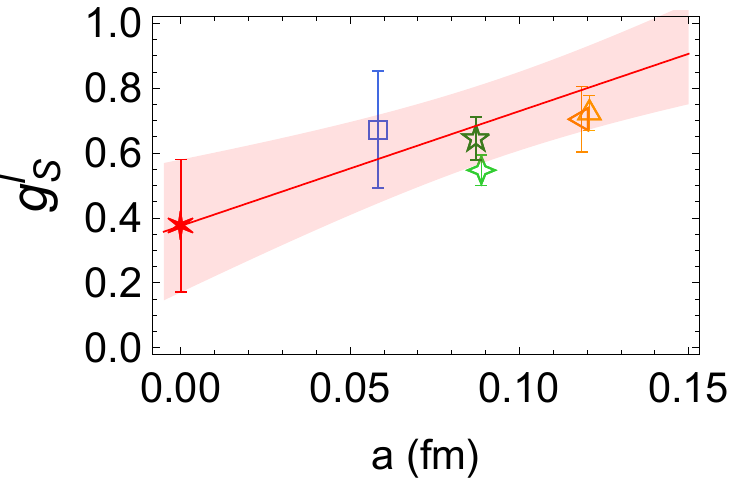}
\includegraphics[height=0.96in,trim={0.86cm  0.02cm 0 0},clip]{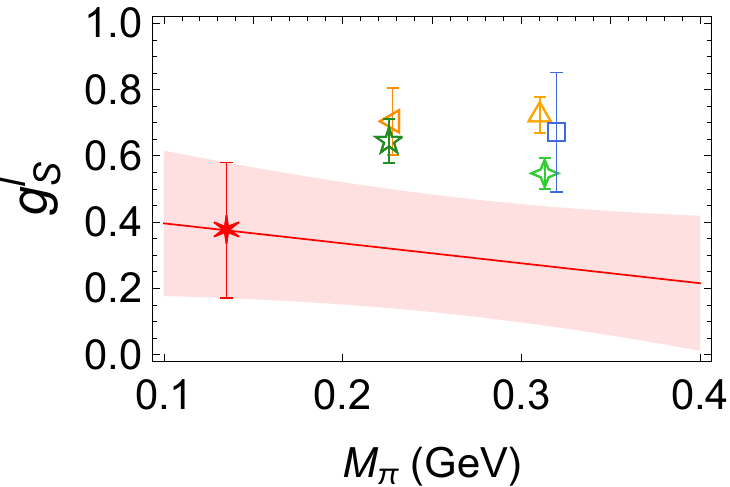}
\includegraphics[height=0.96in,trim={0.05cm  0.02cm 0 0},clip]{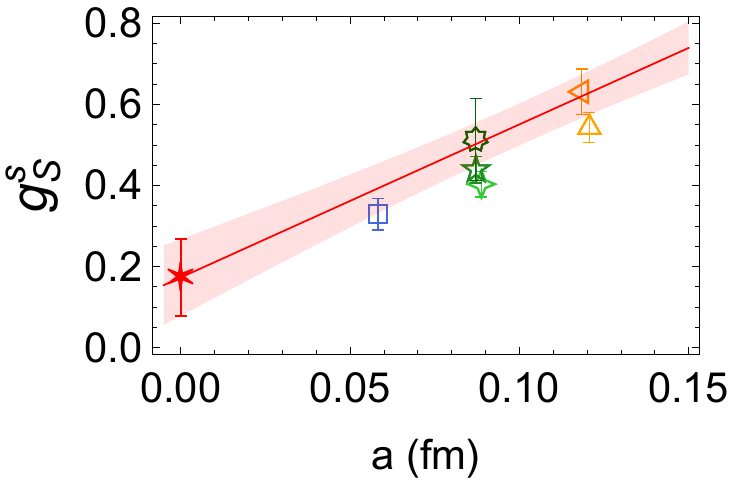}
\includegraphics[height=0.96in,trim={0.86cm  0.02cm 0 0},clip]{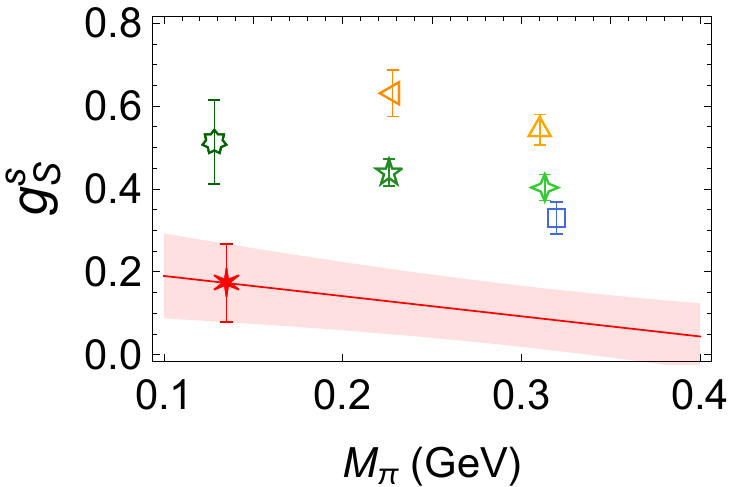}
}
\subfigure{
\includegraphics[height=0.88in,trim={0.05cm  0.02cm 0 0},clip]{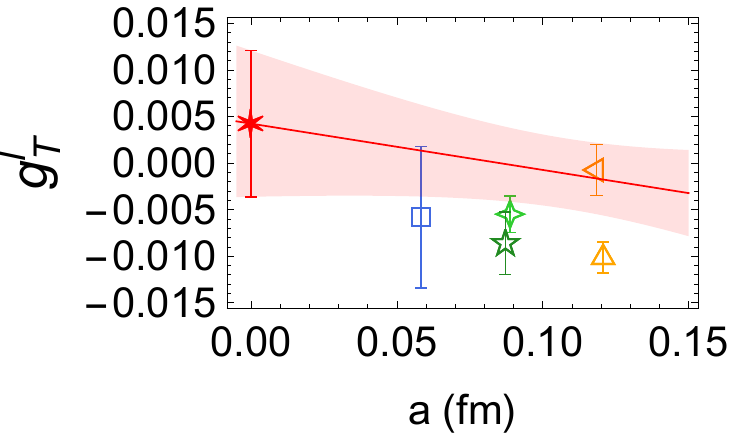}
\includegraphics[height=0.88in,trim={1.00cm  0.02cm 0 0},clip]{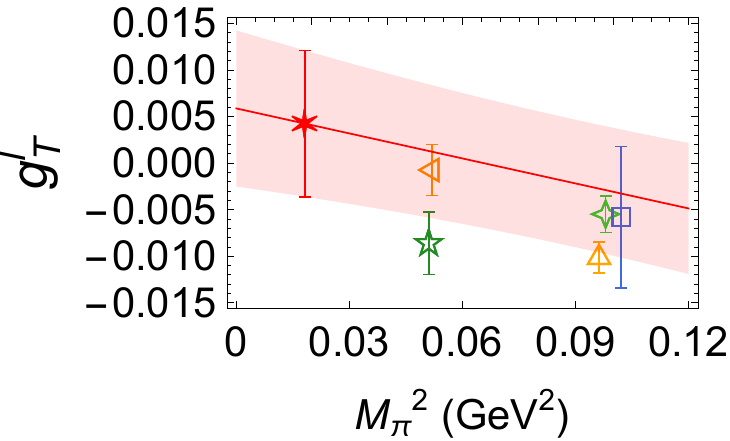}
\includegraphics[height=0.88in,trim={0.05cm  0.02cm 0 0},clip]{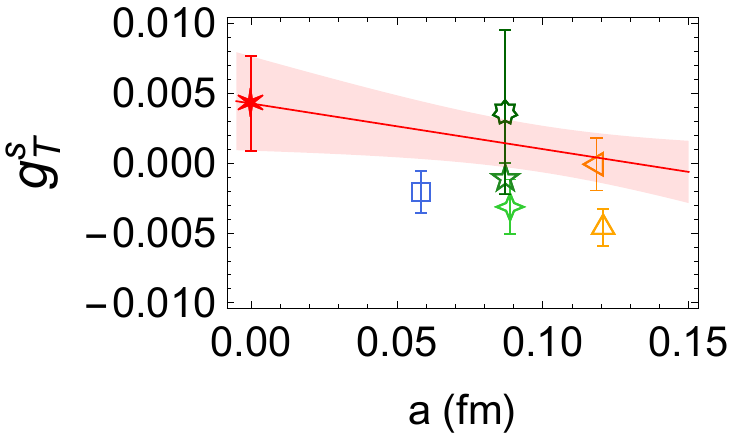}
\includegraphics[height=0.88in,trim={1.00cm  0.02cm 0 0},clip]{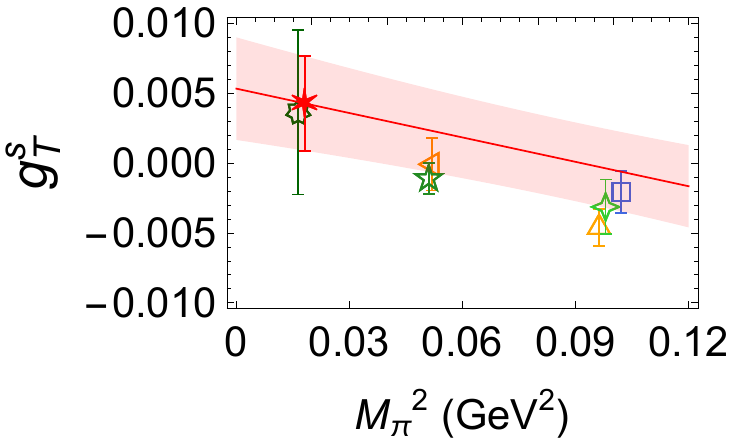}
}
\caption{The continuum-chiral extrapolation for the contributions of
  the disconnected light (left 2 panels) and strange (right 2 panels)
  quarks. Each pair of panels shows the simultaneous fit versus $a$
  and $M_\pi^2$ ($M_\pi$ for $g_S$), and the extrapolated value is
  marked with a red star.  Possible finite volume corrections are
  neglected for lack of sufficient volume dependent data.}
  \label{fig:disc}
\end{figure*}

We have calculated the disconnected contributions of light quarks on 5
ensembles $a12m310$, $a12m220$, $a09m310$, $a09m220$ and
$a06m310$. For the strange quark we added the physical mass ensemble
$a09m130$ and increased the statistics.  The stocastic method used is the same as described in
Ref.~\cite{Bhattacharya:2015wna}. The chiral-continuum plots for these data
are shown in Fig.~\ref{fig:disc}. The renormalization is carried out
using the same factors as for isovector currents. While this has been
shown to be a good approximation for $g_A$ and
$g_T$~\cite{Green:2017keo}, the same is not true for $g_S$. So the
data for $g_S$ in Fig.~\ref{fig:disc} is shown only for completeness.
Our estimates for the axial and tensor charges, after a simultaneous
chiral-continuum extrapolation are:
\begin{align}
  g_A^{l}  &= -0.125(21)             \qquad\qquad\ \        g_A^{s}  = -0.065(12)  \,, \\
  g_T^{l}  &= \phantom{-}0.0042(79)  \qquad\qquad     g_T^{s}  = \phantom{-}0.0043(34) \,.
  \label{eq:discfinal}
\end{align}

Our new result $g_T^{s}=0.0043(34)$ is an improvement over the
previously published value
$g_T^{s}=0.008(9)$~\cite{Bhattacharya:2015wna}.  The result for
$g_T^{l}$ is also still consistent with zero. Based on the current
data, it is reasonable to assume that the magnitude of both after
extrapolation is $\lesssim 0.01$.  Therefore, to get a precise
value will require higher precision data on more ensembles to improve
the chiral-continuum extrapolation in $M_\pi^2$ and $a$.  Given that we can bound
their magnitude to be $\lesssim 0.01$, we will continue to neglect the
disconnected compared to the connected contribution to $g_T^{u}$ and
$g_T^{d}$ as discussed below.

These flavor diagonal tensor charges give the contribution of each
quark's electric dipole moment (qEDM) to the neutron EDM as discussed
in Refs.~\cite{Bhattacharya:2015esa,Bhattacharya:2015wna}.  They are
also probed in the measurements of transversity in deep inelastic
scattering: the tensor charges are the integral over the longitudinal
momentum fraction of the experimentally measured quark transversity
distributions~\cite{Ye:2016prn,Bhattacharya:2015wna}.

Results for the connected parts of the flavor diagonal charges, 
using the same renormalization factor as for the
isovector currents, are 
\begin{align}
  g_A^{u,{\rm conn}}  &= 0.868(25)  \qquad\qquad\      g_A^{d,{\rm conn}}  = -0.331(16)  \,, \\
  g_S^{u,{\rm conn}}  &= 4.78(29)   \qquad\qquad\ \ \  g_S^{d,{\rm conn}}  = \phantom{-}3.71(22)    \,, \\
  g_T^{u,{\rm conn}}  &= 0.806(34)  \qquad\qquad\      g_T^{d,{\rm conn}}  = -0.203(14)    \,.
  \label{eq:FDconnected}
\end{align}
Estimates for all three charges are consistent with
those given in Ref.~\cite{Bhattacharya:2015wna}, and there is no
significant reduction in the errors, which are still dominated by the 
final simultaneous chiral-continuum extrapolation. 

Adding the connected and disconnected contributions for the axial
charges and combining their errors in quadratures because the number of
ensembles analyzed and the statistics are different in the two
calculations, we get
\begin{align}
  g_A^{u} &= 0.743(33) \qquad\qquad g_A^{d} = -0.458(26) \qquad\qquad   g_A^{s} = -0.065(12) \,, \nonumber \\
  g_A^{u,{\rm Expt.}} &= 0.843(12) \qquad\qquad g_A^{d,{\rm Expt.}} = -0.427(12) \,,
  \label{eq:gAuds}
\end{align}
were we also give the experimental
values~\cite{Patrignani:2016xqp}. There is a 2--3$\sigma$ difference
between the lattice and experimental results for both $g_A^{u}$ and
$g_A^{d}$.  The analogous results for the neutron are given by the $u
\leftrightarrow d$ interchange.  From these axial charges, one gets
the contribution of the quarks to the spin of the proton, $\Delta
\Sigma_q/2 = (g_A^{u} + g_A^{d} + g_A^{s})/2 = 0.11(5)$.

%

\section{Summary}

This talk presents the current status of our results for isovector and
flavor diagonal charges of the nucleons using 10 ensembles of
$2+1+1$-flavor HISQ ensembles generated by the MILC
collaboration~\cite{Bazavov:2012xda}. The increase in statistics and
the addition of a second physical mass ensemble has improved the fits,
both to control excited state contamination as well as for the final
chiral-continuum-finite volume extrapolation. Our estimate
$g_A^{u-d}=1.20(3)$ is $2.5\sigma$ below the experimental value. We
find deviations of similar size for the flavor diagonal charges
$g_A^{u}$ and $g_A^{d}$.  Results for the tensor charges are stable
and the error in them is now dominated by the uncertainty in the
renormalization factor. We have corrected an error in the form of the
leading chiral correction used in the final simultaneous fit to the
data for $g_S^{u-d}$, $M_\pi \to M_\pi^2$. As a result, the estimate for
$g_S^{u-d}=1.08(11)$ is about $1\sigma$ larger than the value reported
in Ref.~\cite{Bhattacharya:2015wna}.  Our immediate goal is to double
the statistics on the second physical mass ensemble $a06m135$ and
finalize the analysis for publication.

\vskip20pt

\begin{acknowledgement}
  {\textbf{Acknowledgement}} We thank the MILC Collaboration for
  providing the 2+1+1-flavor HISQ lattices and Emanuele Mereghetti for
  pointing out the correct form of the chiral correction in the
  isovector scalar charge. Simulations were carried out on computer
  facilities of (i) the USQCD Collaboration, which are funded by the
  Office of Science of the U.S. Department of Energy, (ii) the
  National Energy Research Scientific Computing Center, a DOE Office
  of Science User Facility supported by the Office of Science of the
  U.S. Department of Energy under Contract No. DE-AC02-05CH11231;
  (iii) Oak Ridge Leadership Computing Facility at the Oak Ridge
  National Laboratory, which is supported by the Office of Science of
  the U.S.  Department of Energy under Contract No. DE-AC05-
  00OR22725; (iv) Institutional Computing at Los Alamos National
  Laboratory; and (v) the High Performance Computing Center at
  Michigan State University.  The calculations used the Chroma
  software suite~\cite{Edwards:2004sx}. This work is supported by the
  U.S. Department of Energy, Office of Science of High Energy Physics
  under contract number~DE-KA-1401020 and the LANL LDRD program. The
  work of H-W. Lin was supported in part by the M. Hildred Blewett
  Fellowship of the American Physical Society.
\end{acknowledgement}

\bibliography{ref}

\end{document}